\documentclass[showpacs,showkeys,11pt,
preprint,preprintnumbers,nofootinbib,
groupedaddress,superscriptaddress]{revtex4}

\usepackage{amssymb}
\usepackage{amsfonts}
\usepackage{amsmath}
\usepackage{graphicx,subfigure}
\usepackage[export]{adjustbox}
\usepackage{verbatim} 
\usepackage{epsfig}
\usepackage{url}
\usepackage{multirow}
\usepackage{hhline}
\usepackage{feynmp}
\usepackage{bm}
\usepackage{xspace}
\PassOptionsToPackage{naturalnames}{hyperref}
\PassOptionsToPackage{unicode}{hyperref}

\newcommand {\be}{\begin{equation}}
\newcommand {\ee}{\end{equation}}
\newcommand {\ba}{\begin{eqnarray}}
\newcommand {\ea}{\end{eqnarray}}

\usepackage{xcolor}
\usepackage[colorlinks]{hyperref}
\usepackage{comment}
\DeclareUnicodeCharacter{2212}{-}
\begin{document}
\title{Probing Heavy Charged Higgs Boson Using Multivariate Technique at Gamma-Gamma Collider}
\author{Ijaz Ahmed}
\email{ijaz.ahmed@fuuast.edu.pk}
\author{Abdul Quddus}
\email{abdulqudduskakakhail@gmail.com}
\affiliation{Federal Urdu University of Arts, Science and Technology, Islamabad Pakistan}
\author{Jamil Muhammad}
\email{mjamil@konkuk.ac.kr}
\affiliation{Sang-Ho College, and Department of Physics, Konkuk University, Seoul 05029, South Korea}
\author{Muhammad Shoaib}
\email{muhammad.shoaib@iub.edu.pk}
\affiliation{Institute of Physics (IoP), The Islamia University of Bahawalpur, Bahawalpur 63100, Pakistan}
\author{Saba Shafaq}
\email{saba.shafaq@iiu.edu.pk}
\affiliation{International Islamic University Islamabad Pakistan}
\date{\today}

\begin{abstract}
The current study explores the production of charged Higgs particles through photon-photon collisions within the Two Higgs Doublet Model context, including one-loop-level scattering amplitude of Electroweak and  QED radiation. The cross-section has been scanned for plane ($m_{\phi^{0}}, \sqrt{s}$) investigating the process of $\gamma\gamma \rightarrow H^{+}H^{-}$. Three particular numerical scenarios low-$m_{H}$, non-alignment, and short-cascade are employed. Hence using $h^{0}$ for low-$m_{H^{0}}$ and $H^{0}$ for non-alignment and short-cascade scenario, the new experimental and theoretical constraints are applied.
The decay channels for charged Higgs particles are examined in all the scenarios along with the analysis for cross-sections revealing that at low energy it is consistently higher for all scenarios. However, as $\sqrt{s}$ increases, it reaches a peak value at 1$~$TeV for all benchmark scenarios. The branching ratio of the decay channels indicates that for non-alignment, the mode of decay $W^{\pm} h^{0}$ takes control and for short cascade, the prominent decay mode remains $t\Bar{b}$, while in the low-$m_{H}$ the dominant decay channel is of $W^{\pm} h^{0}$.
 In our research, we employ contemporary machine-learning methodologies to investigate the production of high-energy Higgs Bosons within a 3$ $TeV Gamma-Gamma collider. We have used multivariate approaches such as Boosted Decision Trees (BDT), LikelihoodD, and Multilayer Perceptron (MLP) to show the observability of heavy-charged Higgs Bosons versus the most significant Standard Model backgrounds. The purity of the signal efficiency and background rejection are measured for each cut value.
\end{abstract}
\maketitle
\section{Introduction }
A neutral Higgs boson was discovered by the ATLAS and CMS collaborations at the Large Hadron Collider (LHC) in 2012 having approximately a mass of 125 GeV and its properties were consistent with the prediction of Standard Model (SM) Higgs \cite{atlas2012observation,cms2012observation,collaboration2015precise}. Within the SM framework, gauge boson acquire their masses due to the Brout–Englert–Higgs mechanism utilizing the concept of electroweak symmetry breaking (EWSB). The SM of particle physics does not provide any indications of charged Higgs bosons. However, theories beyond the SM propose the existence of charged Higgs bosons and are frequently incorporated into theoretical frameworks such as Two-Higgs-Doublet Models (2HDM), supersymmetric models, composite Higgs models grand unified theories, and axion models. Among all these BSM theories, the two Higgs doublet model is very important due to its structural relevance to many new physics models like MSSM \cite{arhrib2017prospects,ross1975neutral}, composite Higgs models, axion models \cite{veltman1976second,veltman1977limit}. Depending upon the couplings to the quarks, the types of 2HDMs predict different properties and interactions for charged Higgs bosons. A charged Higgs boson would be a more massive counterpart to the SM $W^{\pm}$ and $Z$ bosons, which are carriers of the weak force.
The Higgs sector in 2HDM is extended to incorporate other degrees of freedom that include the prediction of five Higgs candidates of the minimal supersymmetric extension of the SM (MSSM) \cite{atlas2015study,collaborations2015combined}. From these five Higgs boson candidates, two are CP even neutral states $h$, $H$, one is CP odd $A$ state and the remaining two states charge Higgs states $H^\pm$. The discovery of any new scalar Higgs boson either neutral or charged will be a strong hint towards the physics beyond the SM of particle physics and the immediate sign of an extended Higgs sector.\\
The future $e^{+}e^{-}$ and $\gamma\gamma-$ colliders, with high energy and luminosity, offer a great potential for discovering charged Higgs boson. The output rate at a $\gamma\gamma-$ collider could exceed that of $e^{+} e^{-}-$ collisions at the tree level. In 2HDM, the $e^{+} e^{-} \rightarrow H^{+} H^{-}$ process has been analyzed at the tree level, while the $\gamma\gamma \rightarrow H^{+} H^{-}$ process was only studied at the Born level with Yukawa corrections \cite{hashemi2014charged,ma1996yukawa}. Due to the s-channel contribution the $e^{+}e^{-}\rightarrow H^{+}H^{-}$ triumphs over the cross-section, so the rate of production of $\gamma\gamma\rightarrow H^{+}H^{-}$ is higher than the $e^{+}e^{-}\rightarrow H^{+}H^{-}$. The scattering process of $\gamma\gamma\rightarrow H^{+} H^{-}$ has been studied at the one-loop level.\\
This paper will focus on the multivariate analysis of charged Higgs boson production at the photon-photon collider at the International Linear Collider (ILC). Three benchmark points are selected for numerical examination, each with a $\mathcal{CP}$-even scalar mass of $125~GeV$ and couplings consistent with the known Higgs boson. These points are derived from the ``non-alignment'', ``low-$m_{H}$'', and ``short-cascade'' scenarios, and have been accurately delineated within the constraints of current experimental data and are fully consistent with theoretical constraints \cite{haber2016erratum}. The cross-section is scanned for plane $(\phi^{0},\sqrt{s})$, where $\phi^{0}$ is $h^{0}$ for low-$m_{H}$ and $H^{0}$ for non-alignment and short-cascade scenarios. Additionally, the polarization effect is also discussed for all scenarios. 

\section{Review of Two Higgs Doublet Model}
The two scalar doublets are used to acquire masses for gauge bosons and fermions after having their vacuum expectation values (VEVs). The Lagrangian is given by:
\begin{equation}
	\mathcal{L}_{2HDM}= \mathcal{L}_{SM}+\mathcal{L}_{Scalar}+\mathcal{L}_{Yukawa}
\end{equation}
Where $\mathcal{L}_{Scalar}$ is the Lagrangian for two scalar doublets including kinetic term and scalar potential terms. The $Z_{2}$ symmetry is involved to ignore the Flavour Changing Neutral currents (FCNCs), then the transformation for even, $\Phi_{1}\rightarrow+\Phi_{1}$, and for odd is $\Phi_{2}\rightarrow-\Phi_{2}$. To keep $\mathcal{L}_{Yukawa}$ invariant for fermions under $Z_{2}$-symmetry the fermions are coupled with one scalar field:
\begin{equation}  
	\mathcal{L}_{Yukawa}=-\bar{Q}_{L}Y_{u}\tilde{\Phi}_{u}u_{R}-\bar{Q}_{L}Y_{d}\Phi_{d}d_{R}-\bar{L}_{L}Y_{\ell}\Phi_{\ell}\ell_{R}+h.c
 \label{yukawalagrangian}
 \end{equation}
In Equation \ref{yukawalagrangian} the $\Phi_{u,d,\ell}$ is either $ \Phi_{1} $ or $\Phi_{2}$, so based on the discrete symmetry of fermions the 2HDM is classified into four types called as Type-I, II, III, and IV. The review for this relevant study is discussed here for $\mathcal{CP}$-conserving 2HDM. If we assume that in the 2HDM the electromagnetic gauge symmetry is present to perform $SU(2)$ rotation on two doublets for alignment of VEVs of two doublets with $SU(2)$ and the $v=246~GeV$ will occupy one neutral Higgs doublet \cite{craig2013searching}. The two complex doublets, $\Phi_{1}$ from SM and $\Phi_{2}$ from EW symmetry-breaking are used to construct the 2HDM. The scalar potential under $SU(2)_{L}~\otimes~U(1)_{Y}$ invariant gauge group is defined as
\begin{equation}
	\begin{split}
		V_{2HMD} = m^{2}_{1}|\Phi_{1}|^{2}+m^{2}_{2}|\Phi_{2}|^{2}-\biggl[m^{2}_{12}(\Phi^{\dagger}_{1}\Phi_{2})+h.c\biggr]+\dfrac{\lambda_{1}}{2}(\Phi^{\dagger}_{1}\Phi_{2})^{2}+\dfrac{\lambda_{2}}{2}(\Phi^{\dagger}_{2}\Phi_{1})^{2}\\+\lambda_{3}(\Phi^{\dagger}_{1}\Phi_{1})(\Phi^{\dagger}_{2}\Phi_{2})+\lambda_{4}(\Phi^{\dagger}_{1}\Phi_{2})(\Phi^{\dagger}_{2}\Phi_{1})+\biggl[\dfrac{\lambda_{5}}{2}(\Phi^{\dagger}_{1}\Phi_{2})^{2}+\lambda_{6}(\Phi^{\dagger}_{1}\Phi_{1})(\Phi^{\dagger}_{1}\Phi_{2})\\+\lambda_{7}(\Phi^{\dagger}_{1}\Phi_{2})(\Phi^{\dagger}_{2}\Phi_{2})+h.c\biggr]
	\end{split}
 \label{2hdmpotential}
\end{equation}
In the above Equation \ref{2hdmpotential} the quartic coupling parameters are $\lambda_{i}$ $(i=1,2,3...,7)$ and the complex two doublets are $\Phi_{i} ~(i=1,2)$. Hermiticity of the potential forces $\lambda_{1,2,3,4}$ to be real while $\lambda_{5,6,7}$ and $m^{2}_{12}$ can be complex. The Paschos-Glashow-Weinberg theorem suggests that a discrete $Z_{2}$-symmetry can explain certain low-energy observables \cite{glashow1977natural,paschos1977diagonal}. Utilizing this symmetry is crucial to effectively prevent any possibility of FCNCs occurring at the tree level. The $Z_{2}$-symmetry requires that $\lambda_{6}=\lambda_{7}=0$ and also $m^{2}_{12}=0$. If this is not allowed i.e. $m^{2}_{12}$ is non-zero then the $Z_{2}$-symmetry is softly broken for the translation of $\Phi_{1}\rightarrow +\Phi_{1}$ and $\Phi_{2}\rightarrow -\Phi_{2}$. The $Z_{2}$ assignments produce four 2HDM-types as mentioned earlier \cite{branco2012theory,gunion2000print}. Table \ref{2hdmtypes} demonstrates how fermions bind to each Higgs doublet in the permitted kinds when flavor conservation is naturally observed.
\begin{table}[h]
	\centering
	\begin{tabular}{p{80pt} p{80pt} p{80pt} p{10pt}} \hline\hline
		Type& $u_{i}$&$d_{i}$&$\ell_{i}$ \\\hline
		I&$\Phi_{2}$&$\Phi_{2}$&$\Phi_{2}$\\
		II&$\Phi_{2}$&$\Phi_{1}$&$\Phi_{1}$\\
		III&$\Phi_{2}$&$\Phi_{2}$&$\Phi_{1}$\\
		IV&$\Phi_{2}$&$\Phi_{1}$&$\Phi_{2}$\\\hline\hline
	\end{tabular}
	\caption{In 2HDMs with $Z_{2}$-symmetry, Higgs doublets $\Phi_{1}$,$\Phi_{2}$ couple to u-type and d-type quarks, as well as charged leptons.}
	\label{2hdmtypes}
\end{table}

This work focuses only on Type-I and Type-II 2HDM whereas in Type-I only $\Phi_{2}$ doublet interacts with both quarks and lepton similarly as SM. In Type-II the $\Phi_{1}$ couples with d-type quarks and leptons while the $\Phi_{2}$ with only u-type quarks. \\
After electroweak symmetry breaking of $SU(2)_{L}~\otimes~U(1)_{Y}$, the scalar doublet's neutral components gets VEV to be $v_{j}$.
\begin{equation}    
	\Phi_{j}=
	\begin{pmatrix}
		\phi^{+}_{j}\\
		\dfrac{1}{\sqrt{2}}(v_{j}+\rho_{j}+\dot\iota\eta_{j})
	\end{pmatrix}~,~~~~~~~~(j=1,2)
\end{equation}
where $\rho_{j}$ and $\eta_{j}$ are real scalar fields. 
The quartic coupling parameters $\lambda_{1}-\lambda_{5}$ and mass terms $m^{2}_{1}$, $m^{2}_{2}$ are considered as physical masses of $m_{h},m_{H},m_{A},m_{H^{\pm}}$ with $\tan\beta=\dfrac{v_{1}}{v_{2}}$ and mixing term $\sin(\beta-\alpha)$. After $Z_{2}$-symmetry is  broken softly the parameter $m^{2}_{12}$ is given by
\begin{equation}
	m^{2}_{12}= \dfrac{1}{2}\lambda_{5}v^{2}\sin(\beta-\alpha)\cos(\beta-\alpha)=\dfrac{\lambda_{5}}{2\sqrt{2}G_{F}}\biggl(\dfrac{\tan\beta}{1+\tan^{2}\beta}\biggr)
\end{equation}
where the last equality is only for the tree level. By considering $\lambda_{6}$ and $\lambda_{7}$ equal to zero concerning  $Z_{2}$-symmetry, $m^{2}_{12}$, $\tan\beta$ and mixing angle $\alpha$ with four Higgs mass is enough to compute a complete model in physical basis. So, with all this, there are seven independent free parameters to explain the Higgs sector in 2HDM. The terms $m^{2}_{1}$ and $m^{2}_{2}$ are given in the form of other parameters:
\begin{equation}
m^{2}_{1}=m^{2}_{12}\dfrac{v_{2}}{v_{1}}-\dfrac{\lambda_{1}}{2}v^{2}_{1}-\dfrac{1}{2}(\lambda_{3}+\lambda_{4}+\lambda_{5})v^{2}_{2}
\end{equation}
\begin{equation}
	m^{2}_{2}=m^{2}_{12}\dfrac{v_{1}}{v_{2}}-\dfrac{\lambda_{1}}{2}v^{2}_{1}-\dfrac{1}{2}(\lambda_{3}+\lambda_{4}+\lambda_{5})v^{2}_{2}
\end{equation}
The phenomenology is dependent upon the mixing angle with angle $\beta$. In the limit where $\mathcal{CP}$-even Higgs boson $h^{0}$ acts like SM Higgs then it approaches the non-alignment limit which is most favored by experimentalists if $\sin(\beta-\alpha)\rightarrow 1$ or $\cos(\beta-\alpha)\rightarrow 0 $. The $H^{0}$ acts as gauge-phobic such that its coupling with vector bosons $Z/W^{\pm}$ is much more suppressed, but when $\cos(\beta-\alpha)\rightarrow1$ the $H^{0}$ acts SM-like Higgs boson. For the decoupling limits $\cos(\beta-\alpha)=0$ and $m_{H^{0}, A^{0}, H^{\pm}}>>m_{Z}$ so at this limit $h^{0}$ interacts with SM particles completely appears like the couplings of the SM Higgs boson that contain coupling $3h^{0}$.
\section{Constraints from Theory and Experiment}
The theoretical restrictions of potential unitarity, stability, and perturbativity compress the parameter space of the scalar 2HDM potential. The vacuum stability of the 2HDM limits the $V_{2HDM}$. Specifically, $V_{2HDM} \geq 0$ needs to be met for all $\Phi_{1}$ and $\Phi_{2}$
directions. As a result, the following criteria are applied to the parameters $\lambda_{i}$ \cite{deshpande1978pattern,cms2010search}
\begin{equation}	\lambda_{1}>0~~,~~\lambda_{2}>0~~,~~\lambda_{3}+\sqrt{\lambda_{1}\lambda_{2}}+Min(0,\lambda_{4}-|\lambda_{5}|)>0
\end{equation}
Another set of constraints enforces that the perturbative unitarity needs to be fulfilled
for the scattering of longitudinally polarized gauge and Higgs bosons. Besides, the scalar potential needs to be perturbative by demanding that all quartic coefficients satisfy $|\lambda_{1,2,3,4,5} | \leq 8\pi$. The global fit to EW requires $\Delta\rho$ to be $\mathcal{O}(10^{-3} )$ \cite{gfitter2014global}. This prevents
substantial mass splitting between Higgs boson in 2HDM and requires that $m_{H^{\pm}} \approx m_{A} , m_{H}$ or
$m_{h}$. \\
Aside from the theoretical restrictions mentioned above, 2HDMs have been studied in
previous and continuing experiments, such as direct observations at the LHC or indirect B-physics observables. As a consequence, numerous findings have been amassed since then, and the parameter space of the 2HDM is now constrained by all results
obtained. In the Type-I of 2HDM, the following pseudoscalar Higgs mass regions $310 < m_{A} < 410 ~GeV$ for $m_{H} = 150 ~GeV , 335 < m_{A} < 400~GeV$ for $m_{H} = 200~GeV , 350 < m_{A} < 400~GeV$ for $m_{H} = 250~GeV$ with $\tan\beta = 10$ have been excluded by the LHC experiment \cite{aaboud2018search}. Furthermore, the $\mathcal{CP}-$odd Higgs mass is bounded as $m_{A} > 350~GeV$ for $tan\beta < 5$ \cite{aad2015collisions} and the mass range $170 < m_{H} < 360 ~GeV$ with $\tan\beta < 1.5$ is excluded for the Type-I \cite{aad2016measurements}. \\
The $H^{\pm}$ mass is constrained by experiments at the LHC and prior colliders, as well as B-physics observables. The BR($b\rightarrow s\gamma$) measurement limits the charged Higgs mass in Type-II and IV 2HDM with $m_{H^{\pm}}>580~GeV$ for $\tan\beta\geq 1$ \cite{misiak2015updated,misiak2017weak}. On the other hand, the bound is significantly lower in Type-I and III of 2HDM \cite{kanemura2015unitarity}. With $\tan\beta\geq 2$, the $H^{\pm}$ in Type-I and III of 2HDM can be as light as $100 ~GeV$ \cite{enomoto2016flavor,arhrib2017bosonic} while meeting LEP, LHC, and B-physics constraints \cite{aad2015search,khachatryan2015search,aad2013search,aleph2013search,akeroyd2017prospects}.
\section{Benchmark points scenarios}
 We have taken three scenarios \cite{haber2016erratum}: non-alignment, short cascade, and low-$m_{H}$. All of these are taken for $\mathcal{CP}-$even scalar of mass $125~GeV$ and couplings are well arranged with observed Higgs boson. The additional Higgs boson searches leave a considerable portion of their parameter space unconstrained, emphasizing the need for further investigation. Validation of potential stability, perturbativity, and unitarity for each BP was performed using \texttt{2HDMC 1.8.0} \cite{eriksson20102hdmc}.\\
 These benchmark situations, shown in Table \ref{benchmarks}, are created using a hybrid approach, where the input parameters are specified as $(m_{h},m_{H}, \cos(\beta-\alpha), \tan\beta, Z_{4}, Z_{5}, Z_{7})$ with softly broken 2HDM of $Z_{2}-$symmetry, where the $Z_{4,5,7}$ are quartic couplings in Higgs basis of $\mathcal{O}$(1). The mass of charged Higgs and pseudoscalar Higgs in this basis are obtained as:
 \begin{equation}
     m^{2}_{A^{0}}= m^{2}_{H^{0}}s^{2}_{\beta-\alpha}+m^{2}_{h^{0}}c^{2}_{\beta-\alpha}-Z_{5}v^{2}
 \end{equation}
 \begin{equation}
     m^{2}_{H^{\pm}}=m^{2}_{A^{0}}-\dfrac{1}{2}(Z_{4}-Z_{5})v^{2}
 \end{equation}
In the \textbf{non-alignment scenario}, the lightest $\mathcal{CP}-$even scalar $h^{0}$, the discovered Higgs boson is interpreted, with SM-like properties. In an alignment scenario heavy $\mathcal{CP}-$even $H^{0}$
could not decay into GB but in a non-alignment scenario it is allowed by present constraints. In this situation to have a $H^{\pm}$ must satisfy the $b \rightarrow s\gamma$ constraint, and quartic couplings are set to -2. The $\tan\beta$ and $m_{H^{0}}$ are remain free parameters.\\
In the \textbf{short-cascade scenario} the SM like $h^{0}$ is taken exactly to alignment i.e $\cos(\beta-\alpha) = 0$. We considered mass hierarchy such as to decay $H^{0} \rightarrow W^{\pm} H^{\pm}$ or $H^{0} \rightarrow A^{0} Z^{0}$ which results Higgs-to-Higgs decay in small cascade. The $Z_{7}$ and $\tan\beta$ are remained fixed parameters.\\
A \textbf{low}$\bm{-m_{H}}$ \textbf{scenario} is proposed where both $\mathcal{CP}-$even Higgs boson ($h^{0} , H^{0} $) are light. The heavier one is assumed to be an SM-like Higgs boson, resulting in $m_{H^{0}}  = 125 ~GeV$. The heavier $\mathcal{CP}-$even Higgs coupling to gauge bosons is proportional to $\cos(\beta-\alpha)$. Because $m_{h} < m_{H}$, the couplings of lighter $\mathcal{CP}-$even scalars to vector bosons must have been strongly suppressed
to comply with direct search limits, forcing $\sin(\beta-\alpha) = 0$. The parameter space for $90 < m_{h} < 120 ~GeV$ is constrained by searches $h \rightarrow bb, \tau \tau$ at the LHC, which leads to an upper constraint on $\tan\beta$.
\begin{table}[t]
    \centering
    \def\arraystretch{1.4}
    \begin{tabular}{p{50pt} p{50pt} p{50pt} p{40pt} p{40pt} p{30pt} p{30pt} p{30pt}}
    \hline\hline
       \textbf{Scenario} &~~$\mathbf{{m_{h^{0}}}}$\newline\textbf{[GeV]}&~~$\mathbf{{m_{H^{0}}}}$\newline\textbf{[GeV]}&$\mathbf{{c_{\beta-\alpha}}}$&$\mathbf{Z_{4}}$&$\mathbf{Z_{5}}$&$\mathbf{Z_{7}}$&$\mathbf{{t_{\beta}}}$\\ \hline
        \textbf{BP-1} &~~125&150...600&0.1&-2&-2&0&1...20\\ \hline
        \textbf{BP-2}&~~125&250...500&0& -1& 1&-1&2\\ \hline
        \textbf{BP-3}&~~125&250...500&0&2&0&-1&2\\ \hline
        \textbf{BP-4}&65...120&~~125&1.0&-5&-5& 0 &1.5\\ \hline\hline  
    \end{tabular}
    \caption{A set of benchmark scenario input parameters that may be utilized to actualize the 2HDM in Hybrid Basis.}
    \label{benchmarks}
\end{table}
 The mass hierarchy is considered for these benchmark points along with the type of the 2HDMs, shown in Table \ref{mass hierarchy table}. In Table \ref{benchmarks}, $t_{\beta}=\tan\beta$ and $c_{\beta-\alpha}=\cos(\beta-\alpha)$.
\begin{table}[t]
    \centering
    \def\arraystretch{1.4}
    \begin{tabular}{p{90pt} p{50pt} p{80pt} p{90pt}}
    \hline\hline
      \textbf{Scenarios} & \textbf{BP's}&\textbf{2HDM-Type}&\textbf{Mass Hierarchy} \\ \hline
     \textbf{Non-alignment} &BP-1&~~~~~~I&$m_{H^{0}}<m_{H^{\pm}}=m_{A^{0}}$\\ \hline
     \textbf{Short Cascade}&BP-2&~~~~~~I&$m_{A^{0}}<m_{H^{\pm}}=m_{H^{0}}$\\ \hline
     \textbf{Short Cascade}&BP-3&~~~~~~I&$m_{H^{\pm}}<m_{A^{0}}=m_{H^{0}}$\\\hline
     \textbf{Low-}$\mathbf{{m_{H}}}$&BP-4&~~~~~II& $m_{h^{0}}<m_{H^{\pm}}=m_{A^{0}}$\\\hline\hline
    \end{tabular}
    \caption{\label{mass hierarchy table}Mass hierarchy for BP's with 2HDM types used in calculations of cross-section and decay width of charged Higgs.}
\end{table}
\section{The leading order cross-section of charged Higgs production}
Analytical formulations of the cross-section of the $e^{+} e^{-}$ collider for charged Higgs pair generation are presented in this section. The process used in this paper is given as:
\begin{equation}
    \gamma(k_{1},\mu)~~\gamma(k_{2},\nu) \longrightarrow H^{+}(k_{3})~~H^{-}(k_{4})
\end{equation}
where $k_{a}(a=1,....,4)$ represent the four momenta. There are three different diagrams at the tree level that are topologically distinct because of photon coupling as shown in Figure \ref{feyndiagram}. The total Feynman amplitude is given by:

\begin{figure}
\centering     
\subfigure{\includegraphics[width=40mm,height=40mm]{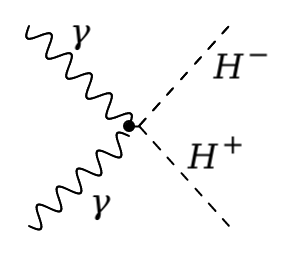}}
\subfigure{\includegraphics[width=50mm, height=40mm]{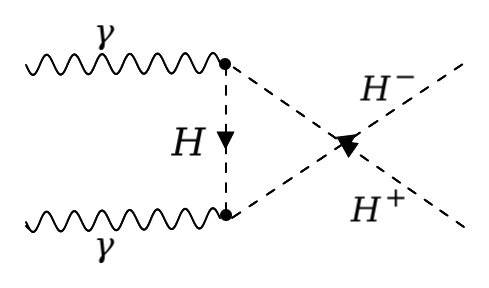}}
\subfigure{\includegraphics[width=50mm,height=40mm]{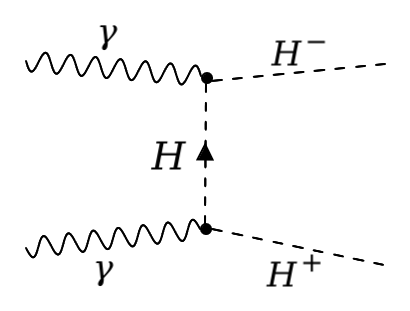}}
\caption{The tree-level Feynman diagrams for the process $\gamma \gamma\rightarrow H^{+}H^{-}$.}
\label{feyndiagram}
\end{figure}

\begin{equation}
\mathcal{M}=\mathcal{M}_{\hat{q}}+\mathcal{M}_{\hat{t}}+\mathcal{M}_{\hat{u}}
\end{equation}
where $\mathcal{M}_{\hat{q}},\mathcal{M}_{\hat{t}}$ and $\mathcal{M}_{\hat{u}}$ are amplitudes of quartic couplings, t-channel, and u-channel Feynman diagrams respectively. The relations for these channels are given as follows:
\begin{equation}  \mathcal{M}_{\hat{q}}=2\dot{\iota}e^{2}g^{\mu\nu}\epsilon_{\mu}(k_{1})\epsilon_{\nu}(k_{2})
    \label{1stfeynmanamplitude}\hspace{4.7cm}
\end{equation}
\begin{equation}
\mathcal{M}_{\Hat{t}}=\dfrac{\dot{\iota}e^{2}}{\Hat{t}-m^{2}_{H^{+}}}(k_{1}-2k_{4})^{\nu}\epsilon_{\nu}(k_{2})(k_{2}+k_{3}-k_{4})^{\mu}\epsilon_{\mu}(k_{1})
    \label{2ndfeynmanamplitude}
\end{equation}
\begin{equation}
    \mathcal{M}_{\Hat{u}}=\dfrac{-\dot{\iota}e^{2}}{\Hat{u}-m^{2}_{H^{+}}}(k_{1}-2k_{4})^{\mu}\epsilon_{\mu}(k_{1})(k_{1}+k_{3}-k_{4})^{\nu}\epsilon_{\nu}(k_{2})
    \label{3rdfeynmanamplitude}
\end{equation}
where the Mandelstam variables are represented by $\Hat{t}=(k_{1}-k_{3})^{2}$ and $\Hat{u}=(k_{2}-k_{4})^{2}$. After calculating the square of the total amplitude and summing up the polarization vectors, the expression becomes:
\begin{equation}
\begin{split}
 |\mathcal{M}|^{2}=256\pi^{2}\alpha^{2}\biggl[\biggl(\dfrac{t}{(t-m^{2}_{H^{+}})^{2}}+\dfrac{u}{(u-m^{2}_{H^{+}})^{2}}-\dfrac{1}{t-m^{2}_{H^{+}}}-\dfrac{1}{u-m^{2}_{H^{+}}}    \biggr)m^{2}_{H^{+}}\\+ \dfrac{9m^{2}_{H^{+}}-3m^{2}_{H^{+}}(2m^{2}_{H^{+}}+s)+(s+t)(s+u)}{2(t-m^{2}_{H^{+}})(u-m^{2}_{H^{+}})} \biggr]
    \end{split}
\end{equation}
The scattering amplitude is calculated numerically in the center of the mass frame, where the four-momentum and scattering angle are indicated by $(k,\theta)$. In the center of mass energy, the energy $(k^{0}_{i})$ and momentum $(\Vec{k}_{i})$ of incoming and outgoing particles are:
\begin{equation}
    k_{1}=\dfrac{\sqrt{s}}{2}(1,0,0,1)~~~~,~~~~k_{2}=\dfrac{\sqrt{s}}{2}(1,0,0,-1)
\end{equation}
\begin{equation}
    k_{3}=(k^{0}_{3},|\Vec{k}|\sin\theta,0,|\Vec{k}|\cos\theta)\hspace{3cm}
\end{equation}
\begin{equation}
    k_{4}=(k^{0}_{4},-|\Vec{k}|\sin\theta,0,-|\Vec{k}|\cos\theta)\hspace{2.5cm}
\end{equation}
\begin{equation}
    k^{0}_{3}=\dfrac{s+m^{2}_{i}-m^{2}_{j}}{2\sqrt{s}}~~~~~~,~~~~k^{0}_{4}=\dfrac{s+m^{2}_{j}-m^{2}_{i}}{2\sqrt{s}}
\end{equation}
\begin{equation}
    |\Vec{k}|=\dfrac{\lambda(s,m^{2}_{H^{+}},m^{2}_{H^{-}})}{\sqrt{s}}\hspace{4cm}
\end{equation}
where $m^{2}_{i}$ is the mass of relevant particles. The cross-section is calculated by taking the flux of incoming particles and the integral over the phase space of outgoing particles is given by:
\begin{equation}
    \Hat{\sigma}_{\gamma\gamma\rightarrow H^{+}H^{-}}(s)= \dfrac{\lambda(s,m^{2}_{H^{+}},m^{2}_{H^{-}})}{16\pi s^{2}}\sum_{pol}|M|^{2}
\end{equation}
In above expression the $\lambda(s,m^{2}_{H^{+}},m^{2}_{H^{-}})$ is the K$\Ddot{a}$llen function relevant to phase space of outgoing $H^{\pm}$. The total integrated cross section for $e^{+}e^{-}$-collider could be calculated by:
\begin{equation}
    \sigma(s)=\int^{x_{max}}_{x_{min}} \Hat{\sigma}_{\gamma\gamma\rightarrow H^{+}H^{-}}(\Hat{s};\Hat{s}=z^{2}s)\dfrac{dL_{\gamma\gamma}}{dz}dz
\end{equation}
where $s$ and $\Hat{s}$ are the C.M enrgy in $e^{+}e^{-}$-collider and subprocess of $\gamma\gamma$, respectively. The value of $x_{min}$ represents the minimum amount of energy needed to generate a pair of charged Higgs particles and is given by $x_{min}=(m_{H^{+}}+m_{H^{-}})/\sqrt{s}$, where the $x_{max}$ is 0.83 \cite{telnov1990problems}. The distribution function of the photon luminosity is:
\begin{equation}
    \dfrac{dL_{\gamma\gamma}}{dz}=2z\int^{x_{max}}_{x_{min}}\dfrac{dx}{x}F_{\gamma/e}(x)F_{\gamma/e}\biggl(\dfrac{z^{2}}{x} \biggr)
\end{equation}
The energy spectrum of Compton back-scattered photons, $F_{\gamma/e}(x)$, is characterized by the electron beam's longitudinal momentum \cite{telnov1990problems}.
\section{NUMERICAL RESULTS AND DISCUSSIONS}
The numerical results of generating charged Higgs boson via photon-photon collisions are thoroughly examined in the context of 2HDM including QED radiations. Cross sections at the tree level are calculated numerically for each benchmark scenario as a function of the C.M energy and the Higgs boson mass. Polarization distributions are presented to improve the production rate by considering longitudinal polarizations of initial beams. Decay pathways of the charged Higgs boson are under study for relevant scenarios.\\
In our work, for analytical and numerical evaluation we have used \texttt{MadGraph5 v3.4.2} \cite{alwall2011madgraph} for the calculations of the cross-sections, the \texttt{2HDMC 1.8.0} \cite{rathsman20112hdmc} for the branching ratio and total decay width. The \texttt{GnuPlot} \cite{williams2012gnuplot} is used for the graphical plotting. 

\begin{figure}
\centering     
\subfigure[]{\includegraphics[width=80mm,height=60mm]{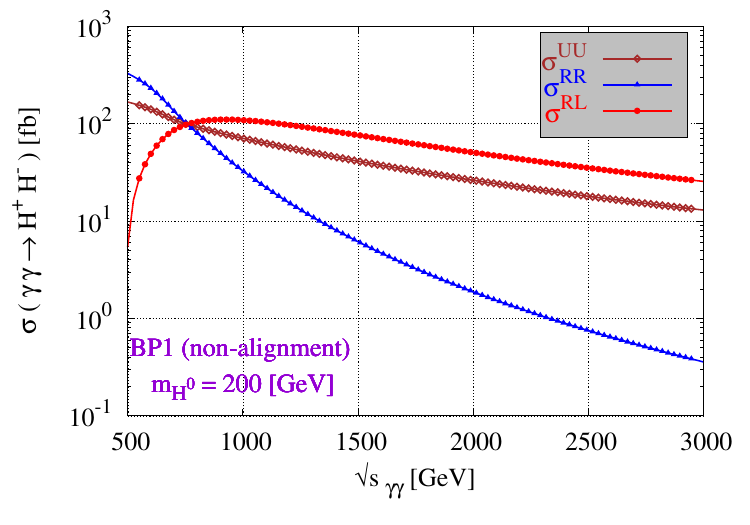}}
\subfigure[]{\includegraphics[width=80mm, height=60mm]{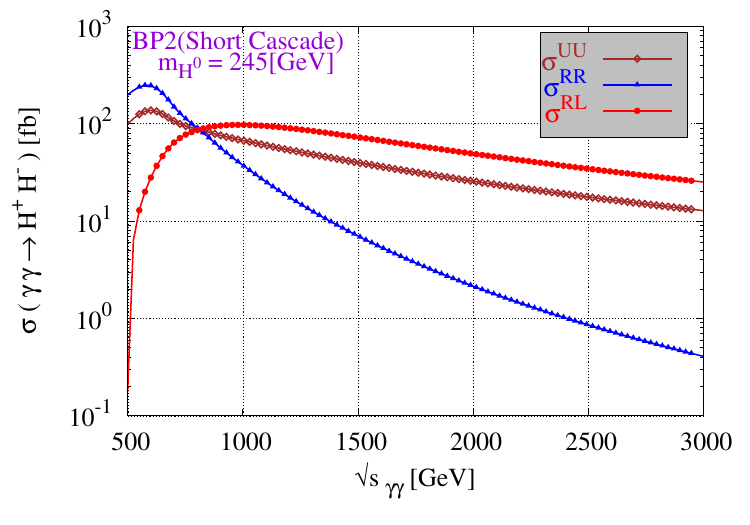}}
\caption{Integrated cross-section for $\gamma\gamma\rightarrow H^{+}H^{-}$ as a function of $\sqrt{s}$ for BP-1 in (a) and BP-2 in (b), respectively}
\label{BP12crosssection}
\end{figure}

As in Figure \ref{BP12crosssection} the cross-section for the process $\gamma\gamma\rightarrow H^{+}H^{-}$ is shown for the C.M energy of $3~TeV$ for three types of polarization; right handed $RR$ ($++$), oppositely-polarized $RL$ ($+-$) and unpolarized beam $UU$.  The cross-section is the same for the polarization modes of $\sigma^{+-}=\sigma^{-+}$. In Figure \ref{BP12crosssection} it can be seen that the cross-section is higher for $UU$ and $RR$ for low $\sqrt{s}$ and gradually decreases. But for the $RL$ mode of polarization, the cross-section reaches a peak value and then gradually decreases. As we can see the cross-section is not enhanced for $RR$ and $UU$ at higher energies but it does only for $RL$. 

\begin{figure}[b]
\centering     
\subfigure[]{\includegraphics[width=80mm,height=60mm]{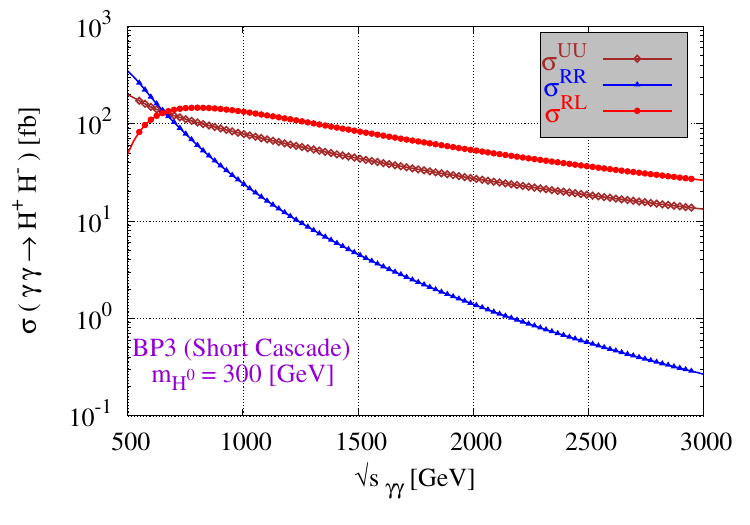}}
\subfigure[]{\includegraphics[width=80mm, height=60mm]{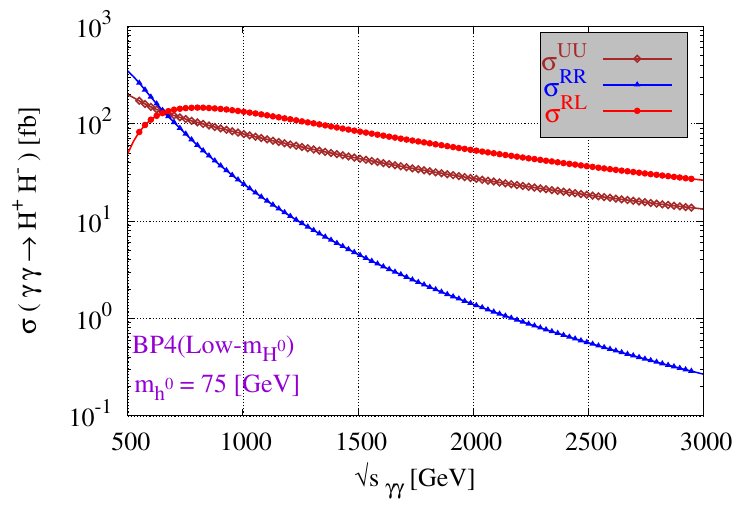}}
\caption{Integrated cross-section for process $\gamma\gamma\rightarrow H^{+}H^{-}$ as a function of $\sqrt{s}$ for BP-3 in (a) and BP-4 in (b), respectively.}
\label{fig:BP34crosssection}
\end{figure}

In Figure \ref{fig:BP34crosssection} for both BPs, the cross-section changes slowly with the mass of $h^{0}$ and $H^{0}$ because of the small range of charged Higgs mass. For both $UU$ and $RR$ modes of polarization the cross-section decreases with C.M energy and for $RL$ mode it reaches a peak value and then decreases. The cross-section $\sigma$  decreases for $\sqrt{s}$ when $m_{H^{\pm}}<<\sqrt{s}/2$. 

\section{decays of charged Higgs Boson}
The probability that a particular particle will decay per unit time is called decay width. While it is impossible to predict the lifespan of a single particle, a statistical distribution can be determined for a large sample, for this purpose the decay width is used. In this section, we will study the final decay products of the charged Higgs bosons created in all scenarios. To investigate the process in a collider, we must first identify all potentially charged Higgs products. The total decay widths of the charged Higgs boson versus the mass of Higgs $h^{0}$ or $H^{0}$ are plotted for all BPs.
\begin{figure}[b]
    \centering
    \includegraphics[width=10cm]{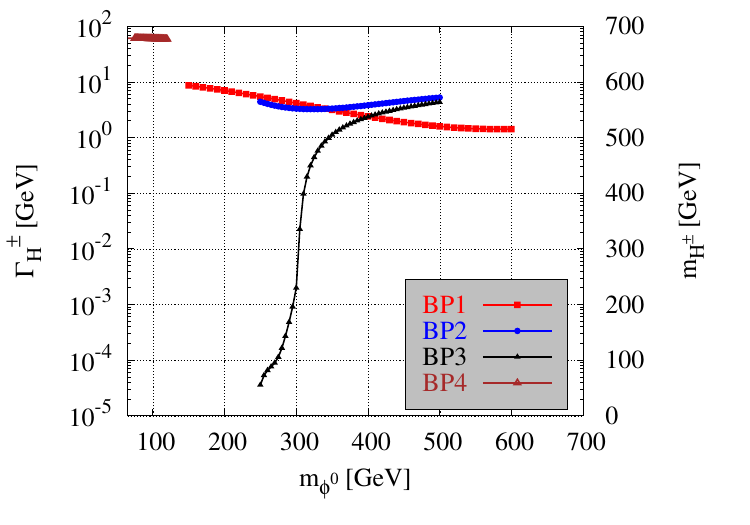}
    \caption{Total decay widths for charged Higgs boson $H^{\pm}$ for all benchmark points.}
    \label{fig:decaywidth}
\end{figure}
The total rate of decay per unit time $\Gamma$ is the sum of all individual decay rates,
$\Gamma=\sum_{j} \Gamma_{j}$. 
 In our system of natural units, the dimension of $\Gamma$ is equivalent to mass (or energy) as it is the inverse of time. As expected, the mass of the charged Higgs boson increases with increasing neutral Higgs mass $m_{\phi^{0}}$ under all circumstances. The decay widths are highly sensitive to the mass hierarchy and mass splitting. Shrinking of the decay width is observed when $m_{H^{0}}-m_{H^{\pm}}$, the mass splitting is minimal, as shown in  Figure \ref{fig:decaywidth}. The decay width for BP-1 decreases from $8.66$ to $1.42$ when the $m_{H^{\pm}}$ goes from $379$ to $691~GeV$. For BP-2, the decay width increases from $4.38$ to $5.22$ for $m_{H^{\pm}}$ in the range of $250 <~m_{H^{\pm}}~<~550~GeV$. The $\Gamma_{H^{\pm}}$ for BP-3 increases from $3.5\times10^{-5}$ to $4.28$ when $m_{H^{\pm}}$ runs from $48.75$ to $436~GeV$. For the last BP-4, the decay width decreases from $60$ to $58$ for change of $m_{H^{\pm}}$ from $558$ to $564~GeV$.\\
 The BR (or branching fraction) is the proportional frequency of a particular decay mode. The BR is the decay rate to the specific mode i.e. $j$  relative to the total decay rate.
\begin{equation}
    BR(j) = \dfrac{\Gamma_{j}}{\Gamma} \nonumber
\end{equation}
 
\begin{figure}
\centering     
\subfigure[]{\includegraphics[width=80mm,height=60mm]{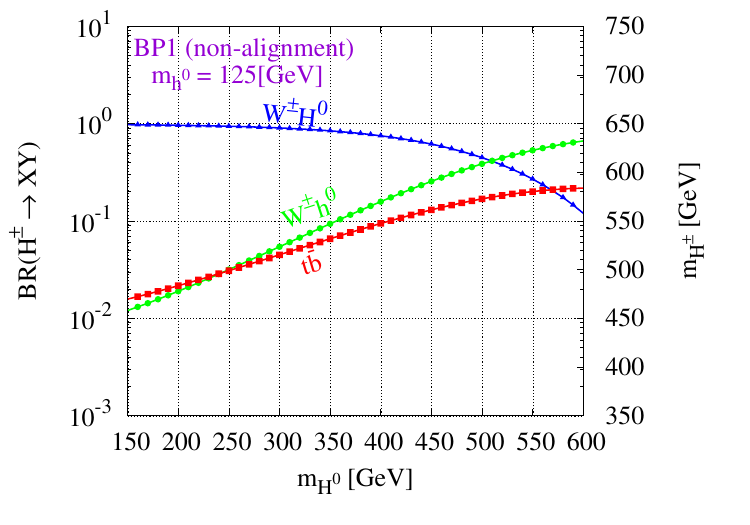}}
\subfigure[]{\includegraphics[width=80mm, height=60mm]{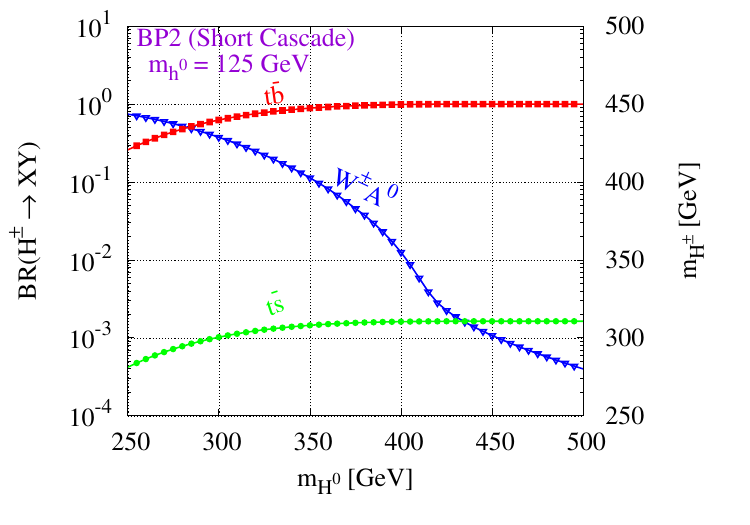}}
\caption{The BR of $H^{\pm}$ predicted in BP-1 from (a) and BP-2 in (b), respectively. For modes, BR is less than $10^{-4}$ are omitted for clarity.}
\label{br12}
\end{figure}
\begin{figure}[b]
\centering     
\subfigure[]{\includegraphics[width=80mm,height=60mm]{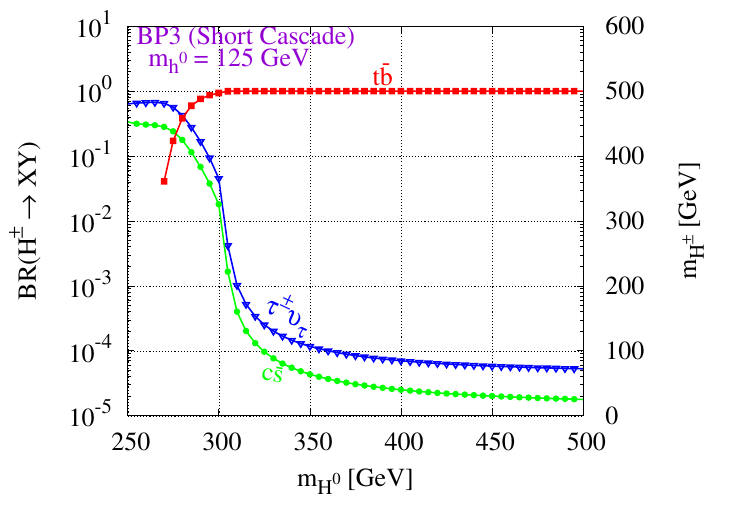}}
\subfigure[]{\includegraphics[width=80mm, height=60mm]{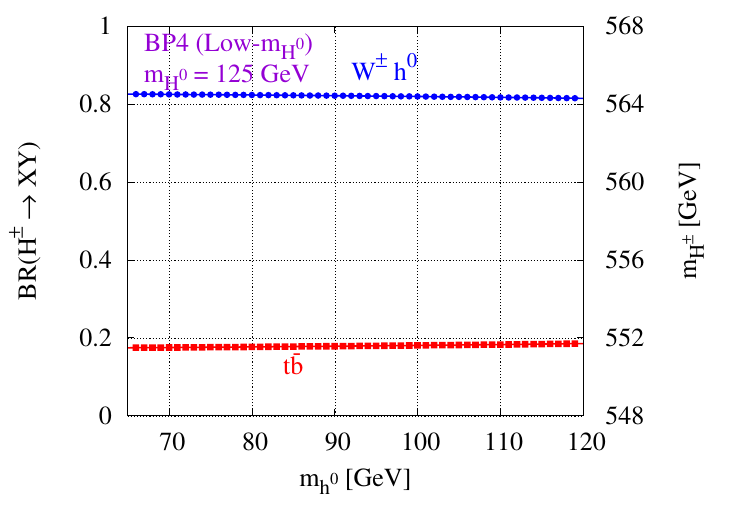}}
\caption{The BR of $H^{\pm}$ predicted in BP-3 from (a) and BP-4 in (b), respectively. For modes, BR is less than $10^{-4}$ are omitted for clarity.}
\label{br34}
\end{figure}
Here $\Gamma$ is the total decay width and $\Gamma_{j}$ is the partial decay width i.e. decay width of an individual particle. We showed the dominant modes of BR for $H^{\pm}$ as function of $h^{0}$ and $H^{0}$, for all scenarios. The  $W^{\pm}H^{0}$ channel is the primary decay mode for $H^{\pm}$ in BP-1, as shown in Figure \ref{br12}. The sub-dominant channels are as follows: $t\Bar{b}$ and $W^{\pm}h^{0}$ for charged Higgs for BP-1,  other suppressed channels are $c\Bar{s}$ and $t\Bar{s}$ for range $m_{H^{0}} < 500~GeV$. The mode of decay $W^{\pm}h^{0}$ takes control when $BR(H^{\pm} \rightarrow W^{\pm}H^{0})$ decreases at larger values of $m_{H^{0}}$. So for $BR(H^{\pm} \rightarrow W^{\pm} h^{0})$ range rises from 1.2 to $66.3\%$ of $150 <m_{H^{0}}<600~GeV$. The process $H^{0}$ to $W^{+}W^{-}$ is also another dominant decay mode with BR of 88.7 to 50.2\% and with hadronic decay of $W^{\pm}$ has 12-jets in the final state.\\
In the BP-2 and BP-3 as shown in Figure \ref{br12} and Figure \ref{br34} respectively, as $m_{H^{\pm}} > m_{t}+m_{b}$ the BR of 100\% prominent decay mode is $H^{\pm} \rightarrow t\Bar{b}$. The suppressed decay modes of $ m_{H^{0}}<300~GeV$ are for $W^{\pm}A^{0}$ and $t\Bar{s}$ in both BP-2 and BP-3; the $t-$quark decay is an ideal for the reconstruction of the process at $m_{H^{0}}>300~GeV$. So for process $t\rightarrow Wb, W\rightarrow q\Bar{q}(l\nu_{l})$ gives $H^{\pm}$ trace at the detector which can be tagged with 8-jets and 2-b-tagged jets.

 For BP-4, shown in Figure \ref{br34}, for a range of $65~<m_{h^{0}}<120~GeV$ the dominant channel is $W^{\pm}h^{0}$ because for $\sin(\beta-\alpha)=0$ it leads to 100\%. 4-jets and 4-b-tagged jets can be used to tag the process.

\section{Multivariate Analysis for charged Higgs production}
An integrated ROOT framework for parallel running and computation of several multivariate categorization algorithms is called the “Toolkit for Multivariate Analysis” \cite{brun1997root}, which categorizes using two sorts of events: signal and background. TMVA especially has many applications in high energy physics for the complex multiparticle final state. To train the classifiers, a set of events with well-defined event types is inserted into the Factory. The event samples for signal and background can either be read using a tree-like structure or a plain text file using a defined structure. All variables that are supposed to separate signal and background events must be known by the Factory. Cuts are applied on signal and background trees separately.\\
We represent three classifiers in our work; Boosted Decision Tree (BDT), LikelihoodD (Decorrelation), and MLP. In BDT a selection Tree is a tree-like structure that illustrates the different outcomes of a choice using a branching mechanism. An event is categorized as either a signal or a background event by passing or failing to pass a condition (cut) on a certain node until a choice is reached. The “root node” of the decision tree is used to find these cuts. The node-splitting process concludes when the BDT algorithm specifies minimal events (\texttt{NEventsMin}). 
\begin{figure}[t]
\centering     
\subfigure[]{\includegraphics[width=80mm,height=60mm]{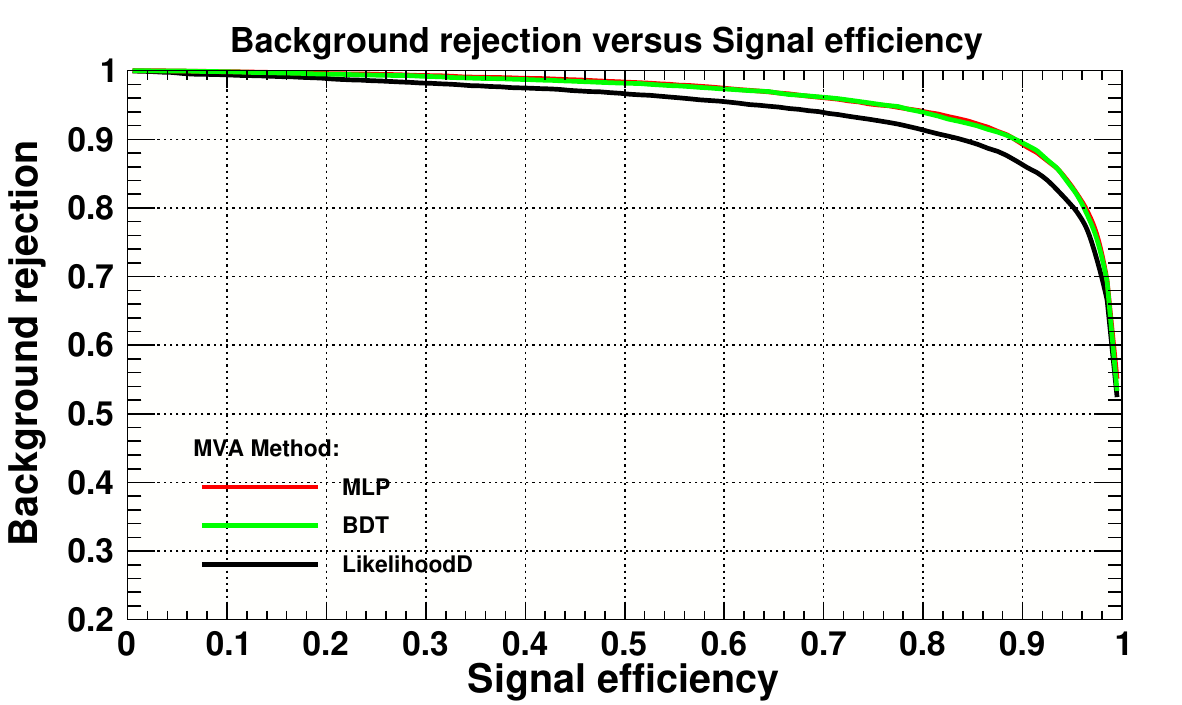}}
\subfigure[]{\includegraphics[width=80mm, height=60mm]{bkg_rjec_sig_eff.pdf}}
\caption{Background Rejection vs Signal Efficiency with applying cuts (a) and without applying cuts (b), respectively.}
\label{bkgrejsigeff}
\end{figure}
\begin{table}[b]
\def\arraystretch{1.5}
    \centering
    \begin{tabular}{p{100pt} p{100pt} c}\hline\hline
        \textbf{MVA Classifier} & \textbf{AUC (with cut)}& \textbf{AUC (without cut)}\\\hline
         \textbf{MLP}& ~~~~~~~0.958 &0.922\\\hline
         \textbf{BDT}&~~~~~~~0.957&0.925\\\hline
         \textbf{LikelihoodD}&~~~~~~~0.941& 0.896\\\hline
    \end{tabular}
    \caption{MVA Classifier Area Under (AUC) the Curve with cuts and without cuts values}
    \label{bkg_rej_sig_eff_table}
\end{table}
The final nodes (leaves) are classified according to their “purity” (p). The value for signal or background (usually $+1$ for signal and 0 or $−1$ for background) depends on whether p is greater than or less than the stated number, e.g. +1 if $p>0.5$ and $-1$ if $p<0.5$ \cite{coadou2022boosted}. To differentiate between the background class and signal, a labeling process is carried out. All occurrences with a classifier output $y > y_{cut}$ are labeled as a signal, while the rest are classified as background. The purity of the signal efficiency $\epsilon_{sig, e f f}$, and background rejection ($1- \epsilon_{bkg, e f f}$) are evaluated for each cut value of $y_{cut}$ \cite{speckmayer2010toolkit}. ADA-Boost algorithm re-weights every misclassified event candidate. The new candidate weight consists of the one used in the former tree multiplied by $\alpha=1-\Delta_{m}/\Delta_{m}$, where $\Delta_{m}$ is the misclassification error. This leads to an increase in the weight and therefore an increase in the candidate’s importance when searching for the best separation values. The weights of each new tree are based on the ones of its predecessor \cite{meir2003introduction}. 
\begin{figure}[t]
\centering     
\subfigure[]{\includegraphics[width=80mm,height=60mm]{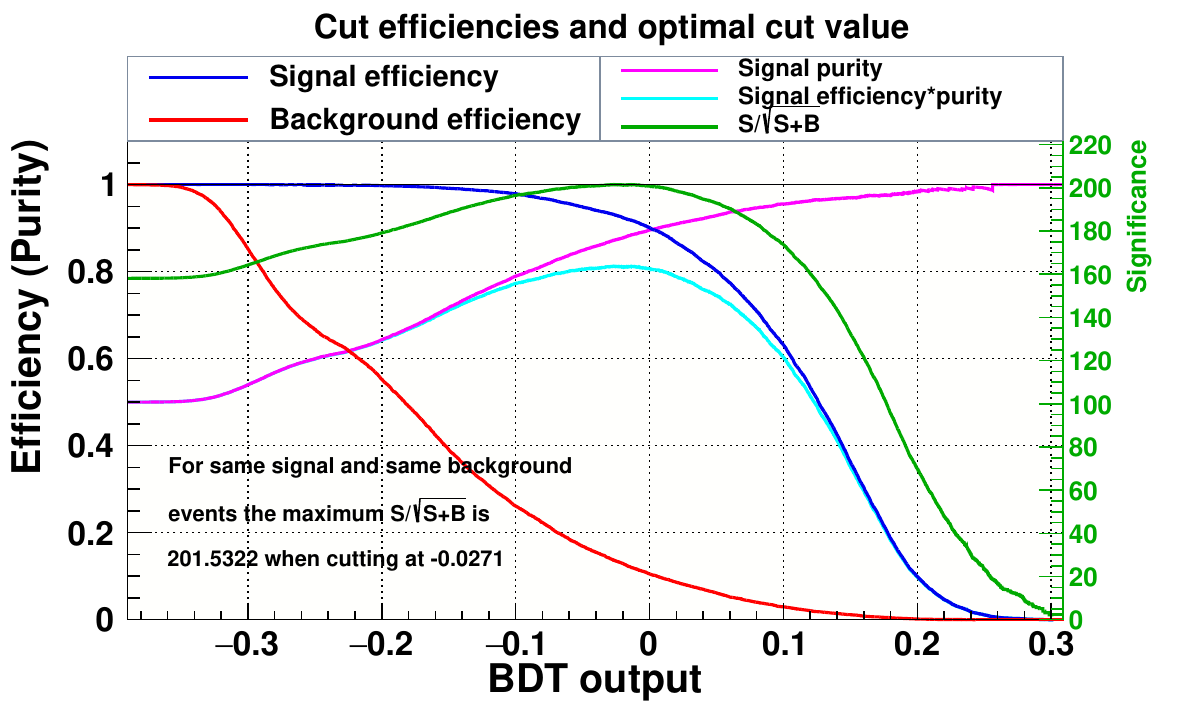}}
\subfigure[]{\includegraphics[width=80mm, height=60mm]{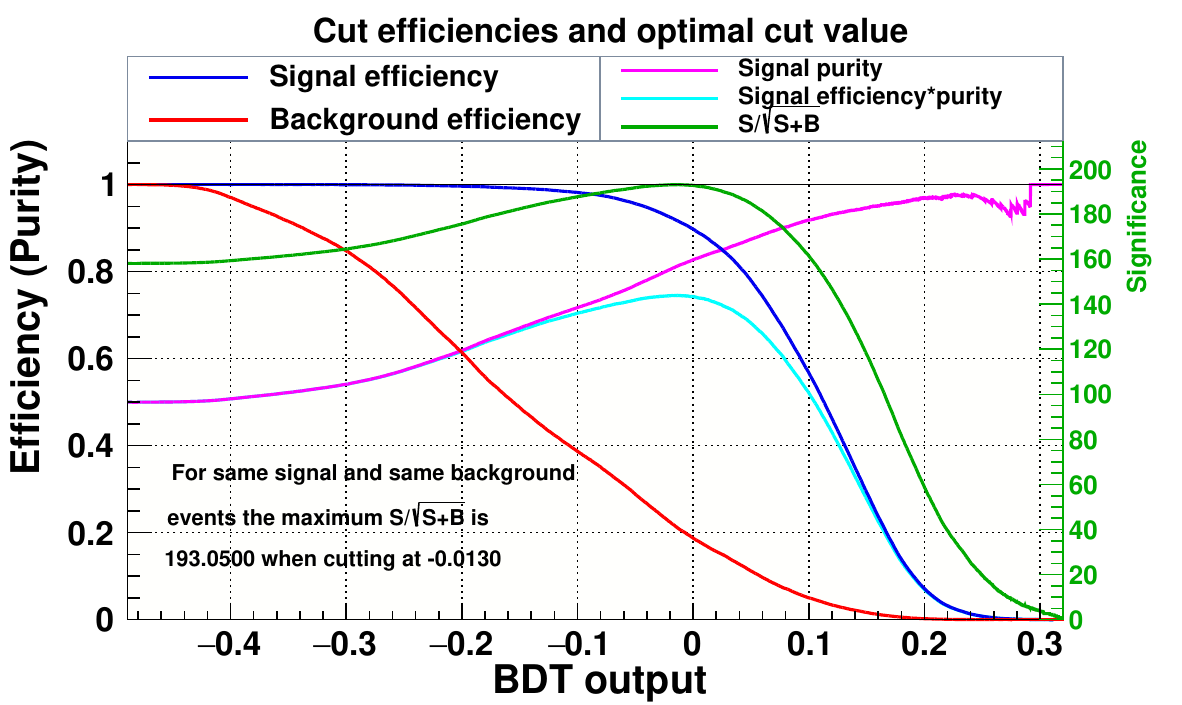}}
\caption{BDT signal significance with applying cuts (a) and without applying cuts (b) respectively.}
\label{cuteffbdt}
\end{figure}
An Artificial Neural Network (ANN) comprises linked neurons, each with its weight. To speed up the processing, a reduced layout can be used as well, the so-called multilayer perception (MLP). The network consists of three kinds of layers. The input layer, consisting of $n_{var}$ neurons and a bias neuron, many deep layers containing a user-specified number of neurons (set in the option \texttt{HiddenLayers}) plus a bias node, and an output layer and each of the connection between two neurons carries a weight.\\
For event $j$, the likelihood ratio $y_{L} (j)$ is defined by
\begin{equation}
    y_{L}(j) = \dfrac{L_{S}(j)}{L_{S}(j) + L_{B}(j)}
    \label{likelihood ratio}
\end{equation}
where the likelihood of a candidate to be signal/background may be determined using the following formula
\begin{equation}
    L_{S/B}(j) =\prod_{i=1}^{n_{var}}P_{S/B,i} (x_{i} (j))
    \label{likelihood S/B}
\end{equation}
where $P_{S/B, i}$ is the PDF for the $i$th input variable
$x_{i}$ . The PDFs are normalized to one for all $i$:
\begin{equation}
     \int_{\infty}^{-\infty}P_{S/B,i} (x_{i})dx_{i} = 1
     \label{PDF normalization}
\end{equation}
The projective likelihood classifier has a major drawback in that it does not use correlation among the discriminating input variables. In the realistic approach, it does not provide an accurate analysis and leads to performance loss. Even other classifiers underperform in the presence of variable correlation.
Linear Correlation was used to quantify the training sample by obtaining the square root of the covariant matrix. The square root of the matrix $C$ is $C^{'}$, which when multiplied by itself yields $C: C=(C^{'})^{2}$. As a result, TMVA employs diagonalization of the (symmetric) covariance matrix provided by:
\begin{equation}
    D = S^{T} CS~~ \implies~~ C^{'} = S\sqrt{ D}S^{T} 
\end{equation}
$D$ is the diagonal matrix, while $S$ denotes the symmetric matrix.
The linear decorrelation is calculated by multiplying the starting variable $\mathbf{x}$ by the inverse of $C^{'}$.
\begin{equation}
    \mathbf{x} ~~\mapsto~~ (C^{'})^{-1}\mathbf{x}
\end{equation}
Only linearly coupled and Gaussian distributed variables have full decorrelation.
In this work, the signal and background events are taken to be 50000 with applied cuts:
\begin{equation}
    P^{Jet}_{T}>30~GeV~~,~~~\eta_{Jet}<2~~~,~~~ N_{Jet}\leq 6~~, ~~~\Delta R < 0.4~~, ~~~E^{Missing}_{T}< 120 ~GeV \nonumber
\end{equation}
\begin{figure}[t]
\centering     
\subfigure[]{\includegraphics[width=80mm,height=60mm]{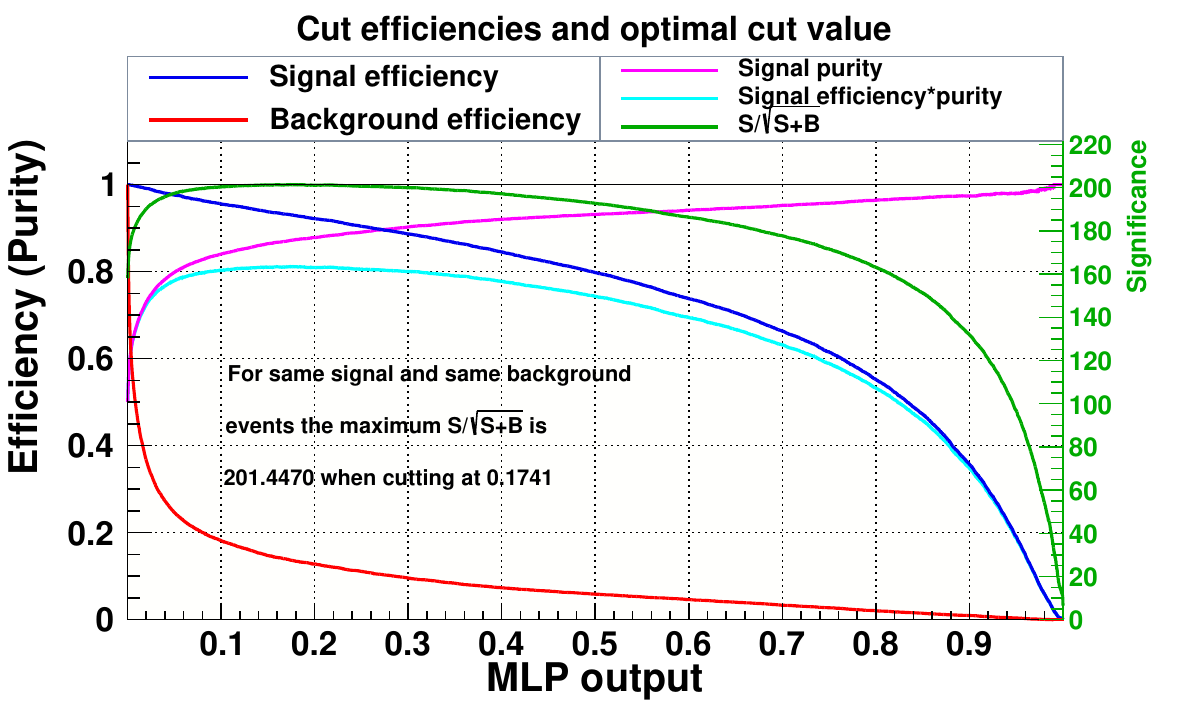}}
\subfigure[]{\includegraphics[width=80mm, height=60mm]{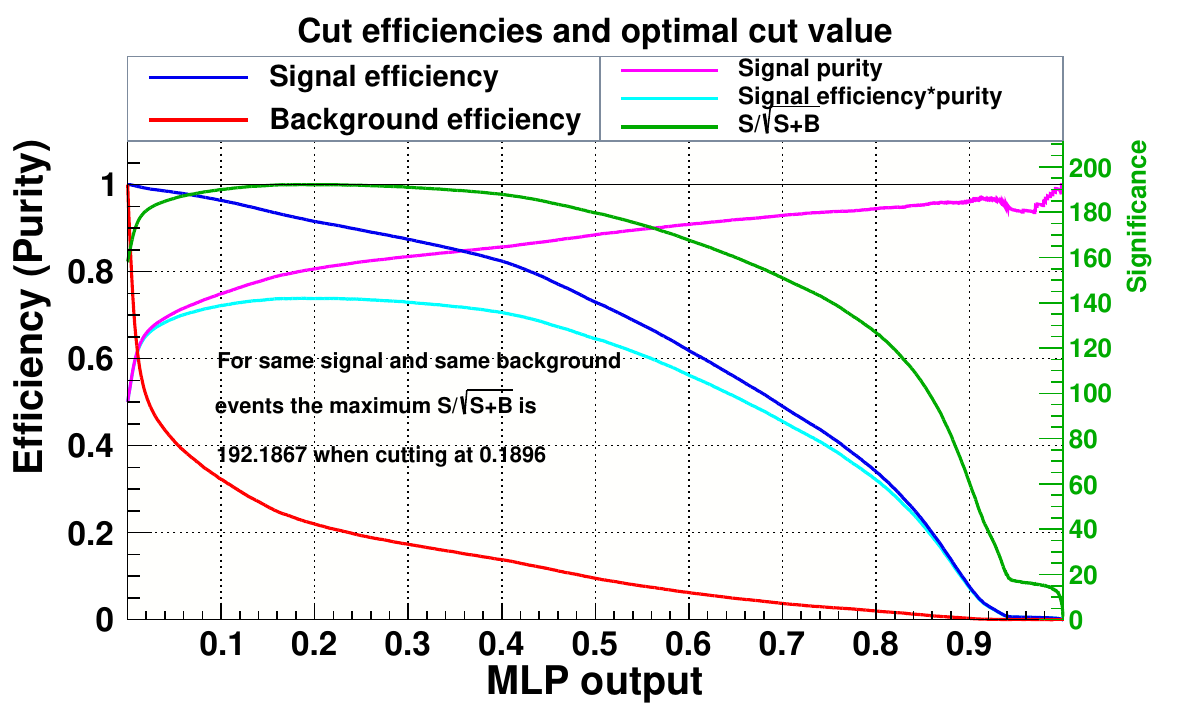}}
\caption{MLP Signal Significance with applying cuts (a) and without applying cuts (b) respectively.}
\label{cuteffmlp}
\end{figure}
\begin{table}[b]
\def\arraystretch{1.5}
    \centering
    \begin{tabular}{p{100pt} p{130pt} p{100pt}}\hline\hline
      \textbf{MVA  Classifier }&\textbf{Signal Significance}\newline\textbf{(with cuts)}&\textbf{Signal Significance}\newline\textbf{(without cuts)}\\\hline
        \textbf{MLP}&~~~~ 201.44 &~~~~192.187 \\\hline 
\textbf{ LikelihoodD}&~~~~198.658 &~~~~188.409 \\\hline
\textbf{ BDT}&~~~~201.532&~~~~193.05 \\\hline
    \end{tabular}
   \caption{The signal significance for the classifiers of signal and background with applied cuts and without applied cuts.}
    \label{signalsig}
\end{table}
The curve of background rejection against signal efficiency provides a reasonable estimate of a classifier’s performance. A classifier’s performance is measured by the area under the signal efficiency versus the background rejection curve, so the bigger the area, the better a classifier’s predicted separation power, as shown in Figure \ref{bkgrejsigeff}. The values of area under the curve (AUC) for Figure \ref{bkgrejsigeff}. Table \ref{bkg_rej_sig_eff_table} shows that the best classifier among all is the MLP and BDT, improved after applying cuts and gave the largest area under the curve.

We used 800 trees to improve the BDT’s performance, with node splitting at $2.5\%$ event threshold. Max tree depth set at 3. Trained using Adaptive Boost with a learning rate of $\beta = 0.5$ parent node and the sum of the indices of the two daughter nodes are compared to optimize the cut value on the variable in a node. For the separation index, we use the \textit{Gini Index}. Finally, the variable’s range is evenly graded into 20 cells. The signal values are taken to be 1 and background values approach to 0. 
\begin{figure}[t]
\centering     
\subfigure[]{\includegraphics[width=80mm,height=60mm]{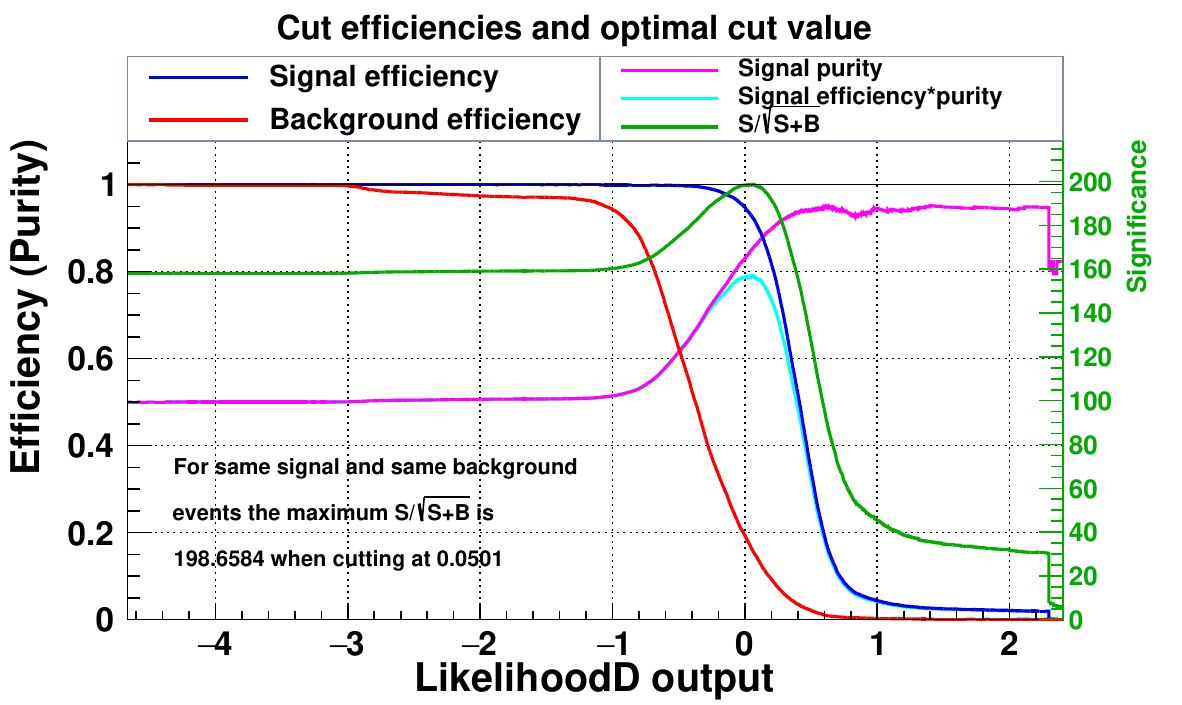}}
\subfigure[]{\includegraphics[width=80mm, height=60mm]{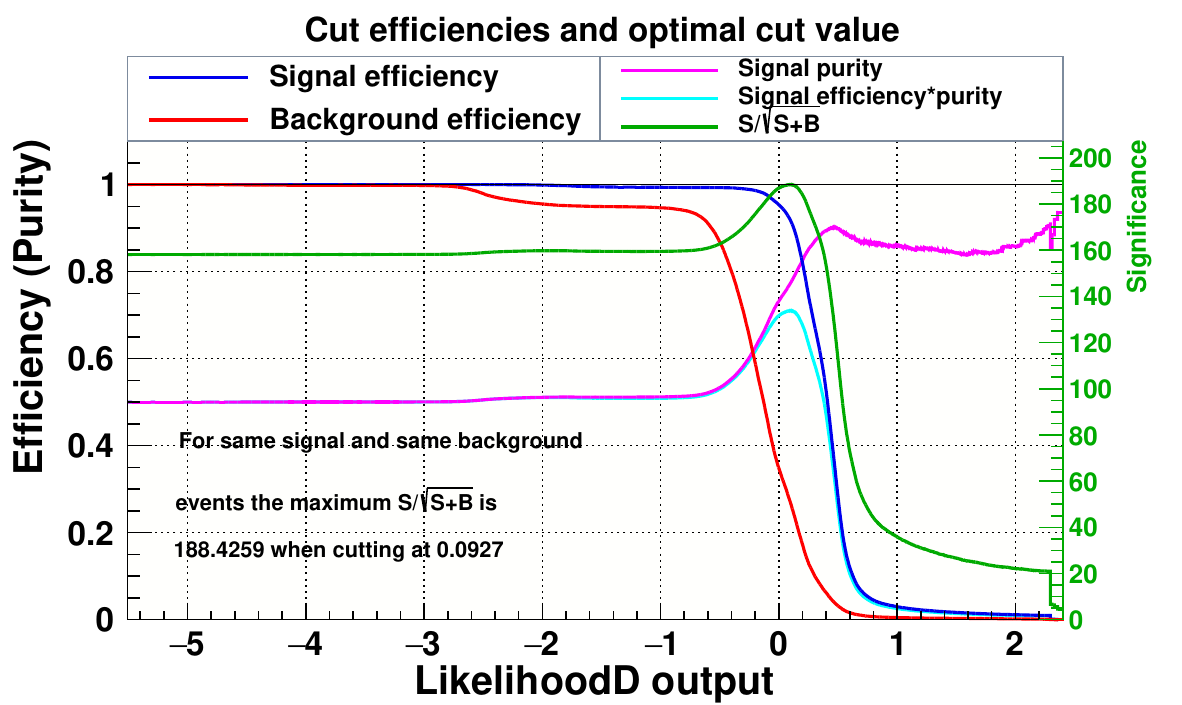}}
\caption{LikelihoodD Signal Significance with applying cuts (a) and without applied cuts (b) respectively.}
\label{cutefflikelihoodd}
\end{figure}
Figure \ref{cuteffbdt} depicts that the signal significance, $S/\sqrt{S+B}$, of the classifier, is improved by applying cuts with an optimal cut of  -0.0271. Similarly, Figure \ref{cuteffmlp} shows the best classifier that is improved by applying cuts with an optimal cut value of 0.1741, and the signal efficiency is also higher than without applied cuts.
 The LikelihoodD signal significance is shown in Figure \ref{cutefflikelihoodd}, has been improved by applied cuts with the optimal cut of 0.0501.
\section{conclusion}

The simplest extension of SM is 2HDM containing charged Higgs Boson and the exact measurement of its nature corresponding model parameters are crucial for the discovery. The pair production is one of the best channels that provided the observable signal in the vast range of parameters in 2HDM. \\
The generation rates of incoming beams are investigated in various polarization collision patterns. The cross-section can be increased twice by oppositely polarized beams of photons at high energies and right-handed polarized beams of photons at low energies as shown in the figures of the cross-section. For BP-1 at the $\sqrt{s}=3~TeV$ the $\sigma^{UU}=12.92\pm 8.1 \times 10^{-06}$ fb , for $\sigma^{RR}=0.3568 \pm 4.6 \times 10^{-07}$ fb and for $\sigma^{RL}=25.48 \pm 1.6 \times 10^{-05}$ fb. For BP-2 the cross section for $\sigma^{UU}=12.76 \pm 8.03 \times 10^{-06}$ fb, $\sigma^{RR}=0.4067 \pm 5.3 \times 10^{-07}$ fb and for $\sigma^{RL}=25.11 \pm 1.8 \times 10^{-05}$ fb at the center of mass energy $3~TeV$. The cross section at $3~TeV$ for BP-3 for different polarizations are $\sigma^{UU}=13.22 \pm 9.1 \times 10^{-06}$ fb, $\sigma^{RR}=0.2674 \pm 3.2 \times 10^{-07}$ fb and $\sigma^{RL}=26.17 \pm 1.7 \times 10^{-05}$ fb respectively. The Low-$m_{H}$ scenario for BP-4 the cross section for polarized beams are $\sigma^{UU}=13.23 \pm 9.1 \times 10^{-06}$ fb, $\sigma^{RR}=0.2673 \pm 3.2 \times 10^{-07}$ fb and for $\sigma^{RL}=26.19 \pm 1.7 \times 10^{-05}$ fb at $\sqrt{s}=3~TeV$ respectively. So we concluded that for all BP's, the cross-section is low at high energy for $UU$ and $RR$ polarized beams of photons, while high for $RL$ at high energy.  \\
For each scenario, the charged Higgs Boson reconstruction has been provided, and its prominent decay modes have been examined. The branching ratio of the decay channel for non-alignment, the bosonic decay channel $H^{\pm}\rightarrow W^{\pm}H^{0}$ is the dominant while in the low$-m_{H}$ scenarios the bosonic decay of $W^{\pm}h^{0}$ is dominant rises to $66.3\%$. The $100\%$ dominant decay channel in a short-cascade scenario is $H^{\pm}\rightarrow t \bar{b}$ that concludes the $t$ decay is the ideal candidate for the reconstruction of the process. Limited phase space and alignment constraints restrict bosonic decay channels.\\
Our Machine Learning models for Multivariate Analysis results are improved by applying cuts. The signal efficiency $(\epsilon_{Sig, eff})$ and background rejection $(1-\epsilon_{Bkg, eff})$ are increased when cuts are applied to the MLP, BDT, and LikelihoodD classifiers. The area under the curve (AUC) is increased for MLP to $3.9\%$, for BDT increased to $3.46\%$, and the LikelihoodD increased up to $5.02\%$ which shows that the LikelihoodD is the more efficient classifier for signal efficiency and background rejection. The signal significance is increased for MLP to $4.81\%$, for the BDT increased to $4.39\%$, and for LikelihooD, it is increased to $5.43\%$ by applying cuts. The significance values obtained with cuts demonstrate how well these models can separate charged Higgs production-related background events from signal occurrences. These cuts most likely aid in lowering background noise, enhancing overall performance, and separating signal events associated with charged Higgs generation. This consistency upholds the validity of the selected machine-learning approaches and increases trust in the outcomes.

\section{Acknowledgements}
We gratefully acknowledge support from the Simons Foundation and member institutions. The current submitted version of the manuscript is available on the arXiv pre-prints home page 
\section{Statements and Declarations}
\textbf{Funding} 
The authors declare that no funds, grants, or other support were received during the preparation of this manuscript.\\
\textbf{Competing Interests}
The authors have no relevant financial or non-financial interests to disclose.\\
\textbf{Availability of data and materials}
Data sharing does not apply to this article as no datasets were generated or analyzed during the current study.

\end{document}